\documentclass[%
superscriptaddress,
 amsmath,amssymb,
 aps,
 twocolumn,
 prl,
 longbibliography
]{revtex4-2}

\usepackage{graphicx}% Include figure files
\usepackage{dcolumn}% Align table columns on decimal point
\usepackage{bm}% bold math
\usepackage{lipsum}
\usepackage{tikz}
\usetikzlibrary{arrows.meta}
\usetikzlibrary{decorations.pathmorphing}
\usepackage{braket}
\usepackage{comment}

\usepackage{hyperref}% add hypertext capabilities
\usepackage{dsfont}
\begin{document}

\pdfoutput=1

\title{Controlling radiative heat flow through cavity electrodynamics}

\author{Francesca Fassioli}
\affiliation{Department of Physics, Friedrich-Alexander-Universit\"at Erlangen-N\"urnberg, D-91058 Erlangen, Germany}

\author{Jerome Faist}
\affiliation{Institute for Quantum Electronics, ETH Zurich, Zurich Switzerland}

\author{Martin Eckstein}
\affiliation{Department of Physics, University Hamburg, D-22607 Hamburg, Germany\\
Email:martin.eckstein@uni-hamburg.de}

\author{Daniele Fausti}
\affiliation{Department of Physics, Friedrich-Alexander-Universit\"at Erlangen-N\"urnberg, D-91058 Erlangen, Germany\\
Email:daniele.fausti@fau.de}
%\email{Email:daniele.fausti@fau.de}

\date{\today}% It is always \today, today,

\begin{abstract}
 
Cavity electrodynamics is emerging as a promising tool to control chemical processes and quantum material properties. In this work we develop a formalism to describe the cavity mediated energy exchange between a material and its electromagnetic environment. We show that coplanar cavities can significantly affect the heat load on the sample if the cavity resonance lies within the frequency region where free-space radiative heat dominates, typically the mid-IR at ambient temperature, while spectral filtering is necessary for having an effect with lower frequency cavities.  

\end{abstract}


\maketitle

Quantum materials (QM) exhibit a remarkable sensitivity to external parameters as underscored by the significant changes in functionalities observed across various QM classes in response to small changes in  temperature and pressure. This susceptibility makes QM an ideal candidate for the optical control of material properties, a possibility that has been extensively explored in time-resolved experiments. 

Studies have already demonstrated the manipulation of metallicity, ferroelectricity, and superconductivity, among others through light-matter interactions, paving the way for further advances in optical control methodologies \cite{Giannetti2016,Fausti2011,delaTorre2021,Murakami2023}.
In this direction, the inherent susceptibility of QM has been at the forefront of the growing research effort focused on modifying material functionalities through the engineering of light-matter interactions by embedding materials into resonant optical cavities
\cite{Schlawin2022}. Numerous theoretical proposals suggest that cavity enhanced vacuum fluctuations 
and the formation of light-matter hybrids
could be the key to control QM properties such as para and ferroelectricity \cite{doi:10.1073/pnas.2105618118,PhysRevResearch.5.043118,PhysRevX.10.041027,Lenk2022}, superconductivity
\cite{Schlawin2019,Sentef2018,Li2020,Curtis2019} magnetic responses 
\cite{Chiocchetta2021,Sentef2020,Fadler2023,Bostroem2023}, as well as topological properties \cite{WangSentef2019,Huebener2021}. 
The concept of integrating and controlling QM with optical cavities is further enriched by the prospect of utilizing ultrashort pulses for Floquet driving schemes in a cavity \cite{Sentef2020,Gao2020,Fadler2023}.

While many of the above theoretical proposals rely on a strong enhancement of vacuum fluctuations in a cavity, which can be challenging to realize, a different path to engineer material properties is found in the context of QM driven by a thermal photon bath.
The presence of an optical cavity can modify the photon exchange between the thermal photon bath and the QM, thereby altering functionalities. In a recent work, we demonstrated the cavity control of a metal to insulator transition in a QM  \cite{Jarc2023}. A  phenomenological estimate suggests that an unrealistically strong light-matter coupling would be needed for the cavity electrodynamics to modify the energy levels through light-matter hybridization. However, we proposed that even under weak light-matter coupling, the cavity can control the radiative heat exchange between the photon environment and the material through a {\it thermal Purcell effect}, which could lead to a change of the material temperature, or more generally its internal state. As originally proposed by Purcell in 1946 \cite{Purcell1946}, and demonstrated decades later by several groups experimentally \cite{Drexhage1974,Vaidyanathan1981,Goy1983,Hulet1985,Gabrielse1985,Martini1987}, transitions can be suppressed (or enhanced) by modifying the electromagnetic environment of a dipole by placing it in an optical cavity.

This paper builds on the {\it Purcell-like model} proposed in \cite{Jarc2023} and introduces a theoretical framework to describe the cavity-mediated exchange of energy between matter and a surrounding photon bath under conditions of weak light-matter coupling. The energy exchange is governed by (i) the local electromagnetic density of states of the cavity, (ii) the linear response of the material, and (iii) the statistical distribution of excitations in the matter and photon fields. The formalism, developed for a general statistical population of photon and matter excitations, is then applied to estimate the cavity-controlled radiative heat load for quasi-thermal matter states in Fabry-Perot cavities. The ensuing discussion addresses the conditions for the non-monotonous behavior of sample temperature with cavity frequency, as observed in \cite{Jarc2023}, shedding light on the intricate interplay between quantum materials and cavity-mediated light-matter interactions. Importantly while here we limit the discussion to the estimate of thermal loads in the weak coupling regime, the formalism could be generalized to the strong coupling regime. Furthermore, it could provide a framework to identify the conditions for the establishment of non-thermal stationary states
\cite{FloresCalderon2023,Curtis2019}, offering a new control mechanism to exploit the unique characteristics of non-thermal states in complex materials.

{\bf Heat flow and local electromagnetic density of states --}
The rate of heat transfer between light and matter, $\dot Q_{lm}$, is given by the expectation value of the Joule heat
\begin{equation}
\dot Q_{lm}=-\int d^3r \langle \hat{\bm J}(\bm r)\hat{\bm E}(\bm r)\rangle ,
\label{joule}
\end{equation}
where $\hat{\bm J}(\bm r)$ is the current density and $\hat{\bm E}(\bm r)$ is the electric field. A detailed derivation is presented in the supplemental material (SM) \cite{suppmat}.  Here we  evaluate Eq. \ref{joule} under the following assumptions: 

(i) The heat flow is dominated by cavity modes which vary weakly on the atomic scale (long wavelength limit). As a consequence, the total $\dot Q_{lm}$ can be computed by a spatial average  $\dot Q_{lm} = \int d^3r \,\dot q_{lm}(\bm r)$, where the {\it heat flow per unit volume $\dot q_{lm}(\bm r)$} at point $\bm r$ in the sample will depend only on the local properties of the material and its environment. 

(ii) The cavity modes are weakly perturbed by the sample (weak light-matter coupling assumption). For given optical properties of the material, the validity of this approximation can be controlled  by the volume fraction of the cavity which is filled by matter. 

(iii) The internal thermalization processes in matter are sufficiently fast, so that matter and photon field are to good approximation in separate thermal equilibrium states at temperatures $T_m$ and $T_l$, respectively. (Technically, field and matter correlation functions approximately satisfy universal fluctuation-dissipation relations with the respective temperatures.)

\begin{figure}
\centerline{\includegraphics[width=0.99\columnwidth]{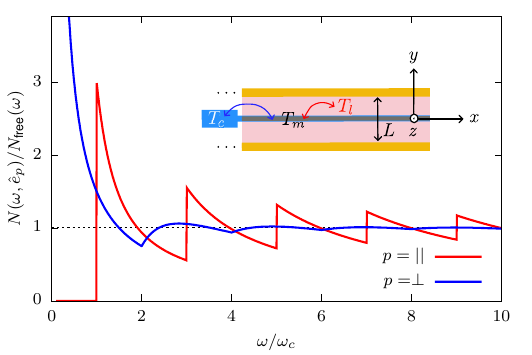}}
\caption{
Local density of states in the median plane of a Fabri-Perot cavity. Ratio $\frac{N(\omega,y=0,\hat e_p) }{N_{\rm free}(\omega)}$ between the local density of states in the coplanar cavity and free space, at the center of the cavity $(y=0)$, for polarization parallel ($p=||$) and perpendicular ($p=\perp$) to the mirrors. Inset: Sketch of the setting, with a sample at temperature $T_m$, connected to a cold bath at temperature $T_c$ and in contact with a photon bath at temperature $T_l$. The spectrum of the latter is affected by the cavity, as exemplified here by two coplanar mirrors. 
}
\label{fig1}
\end{figure}

Under these approximations, $\dot q_{lm}(\bm r)$ can be expressed via the real (dissipative) part of the conductivity $\sigma(\omega,\bm r)$ of the material and the {\it local density of states} of the field $N(\omega,\bm r,\hat e)$.  The latter quantifies the density of modes at frequency $\omega$ in a given polarization $\hat e$, and is in general position dependent. $N(\omega,\bm r,\hat e)$ is directly proportional to the  radiation rate for a point dipole at position $\bm r$ and polarisation $\hat e$ \cite{Barnes2020}.

The scope of this manuscript is to investigate the cavity mediated changes in heat transfer between a photon bath and a material. It is 
therefore convenient to express the heat exchange as a function of the local density of states controlled by the cavity $(N(\omega,\bm r,\hat e_a))$ vs the local density of states in free space $(N_{\rm free}(\omega))$. Denoting by $\hat e_a$ the principle axis of the conductivity tensor, and by $\sigma_{aa}$ the corresponding conductivity, $\dot q_{lm}(\bm r)$ can be written as 
\cite{suppmat}
\begin{align}
\label{Formula}
\dot q_{lm}(\bm r) 
\!=\! 
\sum_{a}  \int_0^\infty \!\!\!\!d\omega
\frac{N(\omega,\bm r,\hat e_a) }{N_{\rm free}(\omega)}
\sigma_{aa}(\omega,\bm r) h(\omega,T_l,T_m),
\end{align}
where
\begin{align}
\label{formulahh}
h(\omega,T_l,T_m)=
\frac{\hbar\omega N_{\rm free}(\omega)}{\epsilon_0}
\Big(
b(\omega,T_l)
-
b(\omega,T_m)
\Big).
\end{align}
Here $b(\omega,T)=1/(e^{\hbar\omega/k_BT}-1)$ is the Bose function, and we have separated out the free space density of states $N_{\rm free}(\omega)= \frac{\omega^2}{3\pi^2c^3}$, such that the free space heat transfer is $\dot q_{lm}^{\rm free} = \int_0^\infty d\omega \, \sum_a \sigma_{aa}(\omega )\,h(\omega,T_l,T_m)$. Equation~\ref{Formula} shows that the heat transfer depends on the local density of states at the frequencies that the material conducts, reflecting that, in principle, the heat flow can be controlled by modifying the electromagnetic environment of the sample by placing it in an optical cavity. 

In this work we  consider coplanar optical cavities such as those used in Ref.~\cite{Jarc2023} (Fig.~\ref{fig1} inset). The local density of states for such a cavity depends on 
the distance to the mirrors ($y$ position in Fig.~\ref{fig1}) and on whether the polarization is parallel $(a=x,z)$ or perpendicular $(a=y)$ to the plates. 
Figure~\ref{fig1}) shows the ratio  $\frac{N(\omega,\bm r,\hat e_a) }{N_{\rm free}(\omega)}$ at the center ($y=0$) of the coplanar cavity  for both parallel  and perpendicular  polarizations (See \cite{Barnes2020} and \onlinecite{suppmat}). Note that $N/N_{\rm free}$ depends on $\omega$  only via the ratio $\omega/\omega_c$, where $\omega_c=c\pi /L$ is the lowest  cavity resonance, with  $L$  the distance between the plates. Crucially, close to the first resonance $\omega=\omega_c$, the density of states can be enhanced or suppressed with respect to that of free space, which is the origin of the Purcell effect. 

In the subsequent sections we will examine the conditions under which the cavity can influence the heat exchange with the photon environment and consequently is able to modify material properties such as temperature. 
In a typical experimental setting (see sketch in Fig.~\ref{fig1}), a material inside an optical cavity is in contact with a cold finger at temperature $T_c$ and the photon bath at temperature $T_{l}$. The temperature of the material $T_m$ is determined by the balance between $\dot Q_{lm}$ and  the heat flow $\dot Q_{mc}$ from matter to the cold reservoir. To significantly influence $T_m$ with respect to the free space limit, it is therefore essential for the cavity to induce a substantial change in the ratio
\begin{align}
\label{eqR}
R= \dot q_{lm}\,/\,\dot q_{lm}^{\rm free}.
\end{align} 

Below,
we will therefore examine the ratio \eqref{eqR} in various 
limits, for the case of a coplanar cavity.

\begin{figure}
\centerline{\includegraphics[width=0.99\columnwidth]{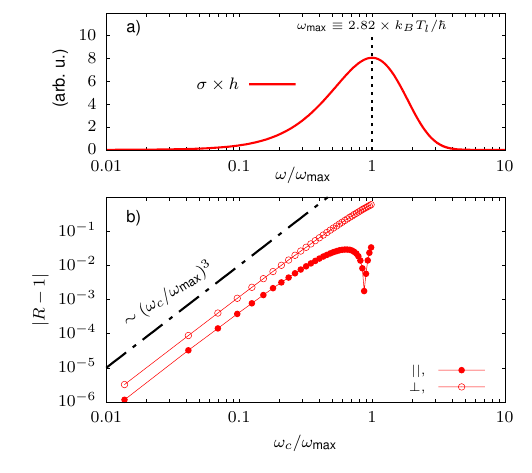}}
\caption{
Heat transfer in low frequency cavities. a) Spectral density for the heat transfer [integrand $\sigma(\omega) h(\omega,T_l)$, Eq.~\eqref{formulahh}] at $T_m=0$ in free space, for a frequency independent $\sigma$; the peak position at $\omega_{\rm max}\approx 2.82 k_BT_l/\hbar$ is taken as an energy scale; e.g., for $T_l=300$K, $\omega_{\rm max}/2\pi\approx6.25\times2.82\approx 17.6$THz. b) The ratio \eqref{eqR} as function of the cavity frequency $\omega_c$, for conductivity purely in plane ($||$), out of plane ($\perp$). For $\omega_c\ll\omega_{\rm max}$, the ratio drops like $(\omega_c/\omega_{\rm max})^3$ or faster (see dash-dotted line).
}
\label{fig2}
\end{figure}

{\bf Heat flow in low-frequency cavities -- } 
First of all, it is important to recognise that the density of states in free space increases with frequency ($N_{\rm free}(\omega)\propto \omega^2$), which is also the overall trend for the density of states in the coplanar cavity. This means that the integrand in Eq.~\ref{Formula} is often dominated by the higher-frequency end of the allowed transitions  in the material. To illustrate this point, consider $\sigma$ to be isotropic and frequency independent. In this case, the maximum of the integrand for the free space scenario is controlled by temperature and the heat flow can be evaluated as 
\begin{equation} 
\dot q_{lm}^{\rm free}(\bm r)=\frac{\sigma\pi^2 k_B^4}{15\epsilon_0c^3\hbar^3}\big(T_l^4-T_m^4\big),
\label{T4}
\end{equation}
corresponding to the Stefan-Boltzmann law. Note that a linearized  form of the heat transfer, 
$\dot q_{lm} \propto T_l-T_m$, 
which was assumed in the analysis of Ref.~\onlinecite{Jarc2023}, is obtained from Eq.~\eqref{Formula}
in the opposite limit of the heat transfer being dominated by low frequencies $\hbar \omega \ll k_BT_l,k_BT_m$, so that  $b(\omega,T)\approx k_BT/\hbar\omega$ in Eq.~\eqref{formulahh}.

Equation~\ref{T4} suggests that placing a material with a flat spectral density in an optical cavity with a low fundamental frequency ($\omega_c\ll k_BT/\hbar$), will not significantly modify the heat exchange with respect to free space, since the frequency range where the cavity density of states differs significantly to that of free space ($\omega\sim\omega_c$) contributes little to the overall heat load. Indeed, let us consider the heat flow in free space and in a cavity with $\omega_c \ll k_BT_l/\hbar$ 
($T_m=0$) for 
a flat $\sigma=constant$ conductivity. Figure~\ref{fig2}a) shows the function  $\sigma(\omega)h(\omega,T_l,T_m)$ which determines the heat transfer in free space (Eq.~\eqref{formulahh}),
peaked at  $\omega_{\rm max}\approx 2.82 k_BT/\hbar$. 
For  frequencies $\omega_c  \ll \omega_{\rm max}$, one finds that the ratio R between the heat flow in the cavity and in free space [Eq.~\eqref{eqR}] approaches $1$ like $R-1= \mathcal{O}((\omega_c/\Omega)^3)$. This is seen in Fig.~\ref{fig2}b), which plots $|R-1|$.
An analytical analysis of the free space limit is given in the  SM \cite{suppmat}. 
  This implies the somewhat intuitive but important result that in order for the thermal Purcell effect to be relevant, the cavity frequency must be in a frequency range which has a significant contribution to the free space heat transfer.  At room temperature ($T_l=300$K, $\omega_{\rm max}/2\pi\approx6.25\times2.82\approx17.6$THz), cavities with frequencies $\omega_c < 1$THz, as for the experiment in \cite{Jarc2023}, would have almost no effect on the heat transfer, unless there is an additional low frequency contribution to $\sigma$ (see discussion section).

\begin{figure}
\includegraphics[width=0.99\columnwidth]{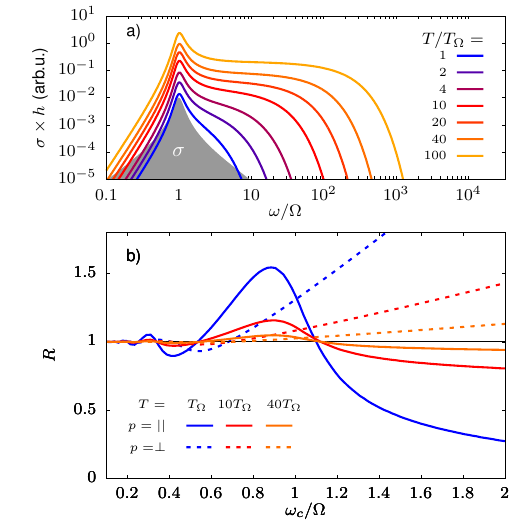}
\caption{
Thermal Purcell effect for a single Lorenz oscillator $(\gamma=0.3\Omega)$.
a) Spectral density for the heat transfer [Eq.~\eqref{formulahh}] for $T_m=0$ and various temperatures $T_l$ in units of $T_\Omega=\hbar\Omega/k_B$ (e.g., for $\Omega/2\pi=1$THz$, T_\Omega=48$K and $T_l\approx 6 T_\Omega$ for $T_l=300$K ). b) The ratio \eqref{eqR} for $T_l=T_\Omega$ (blue lines), $T_l=10T_\Omega$ (red lines),  and $T_l=40T_\Omega$ (orange lines),  as a function of $\omega_c$; solid (dashed) lines are for conductivity purely in plane (out of plane). 
}
\label{fig3}
\end{figure}

{\bf Thermal Purcell effect close to resonance --}
Next we discuss the conditions under which the thermal Purcell effect becomes significant. For illustration, we take a single Lorentz oscillator,
\begin{align}
\label{lorenz}
\sigma_{\Omega,\gamma,\omega_p}(\omega) &=  \epsilon_0\frac{\omega_p^2 \gamma \omega^2}{(\Omega^2-\omega^2)^2 + \omega^2 \gamma^2},
\end{align}
with frequency $\Omega$ and damping constant $\gamma$, and keep the material at zero temperature ($T_m =0$). Figure \ref{fig3}a again shows  the corresponding  integrand for the heat transfer in free space, Eq.~\eqref{formulahh}, for different temperatures $T_l$. 

If $k_BT_l$ is comparable to $\hbar\Omega$, the heat transfer is dominated by the frequency range of the oscillator.  
In this case, varying $\omega_c$ in the vicinity of $\Omega$  and above can be used to either enhance or suppress the heat flow  (see curves for $T=T_\Omega\equiv\hbar\Omega/k_B $ in Fig.~\ref{fig3}b). This non-monotonous behaviour is in agreement with the non-monotonous temperature behavior with cavity length observed in Ref.~\cite{Jarc2023}.
\cite{footnote_giuliano}
 For $\omega_c>\Omega$ one can also achieve a strong anisotropy, as $\dot q_{lm}$ is enhanced  if the conductivity is mainly perpendicular to the plates, and decreased for an in-plane conductivity.  For $k_BT_l\gg \hbar\Omega$, on the other hand, the increase in the density of states with $\omega$ implies that  the heat transfer is dominated by the high frequency tails of the oscillator (Fig.~\ref{fig3}a), up to $\hbar \omega\sim k_BT$. In line with the discussion above, this implies that a cavity $\omega_c\sim\Omega$ would have little effect on the heat transfer,
 as evident from the ratio $R$ for a temperature $T=40T_\Omega$ shown in Fig.~\ref{fig3}b). We stress however that this analysis is based on a lorentzian form for the conductivity and other models with sharper absorption lines (e.g. exponential, see appendix discussion for   1T-TaS$_2$)  
 would relax the condition for temperature.

\begin{figure}
\centerline{\includegraphics[width=0.99\columnwidth]{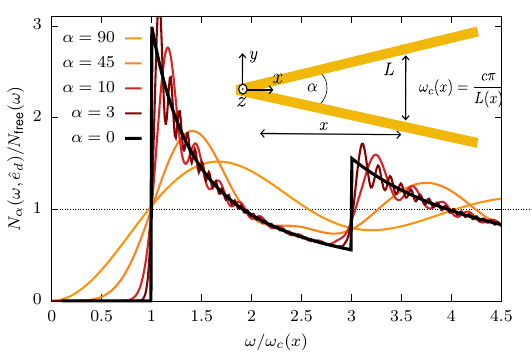}}
\caption{
Local density of states in a misaligned Fabry Perot cavity. Ratio $N_{\alpha}(\omega,\bm r,\hat e)/N_{\rm free}(\omega)$ at a point $\bm r=(x,0,0)$ in a tilted cavity (see inset) for a sketch of the setup); $N/N_{\rm free}$ depends only on $\omega/\omega_c$, where $\omega_c(\bm r) = c\pi/L(\bm r)$ is the local resonance frequency at $\bm r$. For simplicity, results are only shown for the polarization $\hat e=\hat x$.
}
\label{fig4}
\end{figure}

{\bf Tilting the cavity mirrors --} As proposed in Ref.~\cite{Jarc2023}, an indirect evidence for the thermal Purcell effect can be that $T_m$ is modified  not only by changing the cavity frequency, but also by other cavity parameters, such as the quality factor, or the alignment of the cavity mirrors. While a change of the cavity quality factor just washes out the density of states and brings the setting closer to free space, a tilting of the cavity mirrors cannot be translated into a change of the quality factor.
We consider two perfect metallic plates (mirrors), with an opening angle $\alpha$ (see inset in Fig.~\ref{fig4}). We choose the orientation such that the plates are spanned by vectors $\hat e_z$ and $\cos(\alpha/2)\hat e_x \pm  \sin(\alpha/2)\hat e_y$ and intersect at $x=y=0$. If the local density of states is measured at a point $(a,0,0)$, we can define a local cavity frequency $\omega_c(a)=\pi c /L(a)$, $L(a)=2a\tan(\alpha)$. Scale invariance of the Maxwell equations implies that the ratio $N_\alpha(\omega,a,\hat e)/N_{\rm free}(\omega)\equiv \eta(\frac{\omega}{\omega_c(a)},\alpha,\hat e)$ depends only on the dipole orientation, the angle $\alpha$, and finally on the ratio $\omega/\omega_c(a)$ at position $a$.  If the internal relaxation processes are fast enough to establish a roughly uniform temperature in spite of the inhomogeneous coupling to the photon bath, the total heat transfer per volume is obtained by replacing  $N/N_{\rm free}$ in Eq.~\eqref{Formula} with the corresponding spatial average.
The function $\eta$ can be obtained almost analytically (using the method of image dipoles) for angles $\alpha=\pi/N$ (see SM \cite{suppmat}). Figure~\ref{fig4} shows that $ \eta(\omega/\omega_c,\alpha,\hat e)$ is in fact relatively weakly dependent on $\alpha$, and approaches the result for parallel plates already for relatively large opening angles $\alpha$. 
For small angles, we can thus approximate $\eta(\alpha)\approx \eta(\alpha=0)$, so that the  heat flow changes  mainly due to an averaging over different resonant frequencies over the extent of the sample, rather than due to a change in the spectral shape of $N$.

{\bf Discussion and Conclusion --} 
The major result of our analysis is that a necessary condition for the cavity control of the heat flow, both in magnitude and in terms of anisotropy,  is that the cavity resonance is close to the frequency range which dominates the heat transfer outside of the cavity and not only where the conductivity peaks. Indeed, the greatest contribution to heat flow will often be determined by  the high energy tail in the statistical distribution of excitation where the electromagnetic density of states is greater, which is controlled by the temperature in thermal distributions. 

Consequently, for controlling heat flow through a resonance in conductivity at $\Omega$,  the condition $\omega_c\gtrsim\Omega$ should be satisfied. Furthermore, for broad optical conductivity features, a second necessary condition is that the photon temperature satisfies   $T_l\lesssim\Omega$. The calculation indicates that in the majority of the cases for QM with a photon bath fixed at ambient condition ($T=300$K), an appreciable cavity mediated heat load could be observed only for cavities tuned on mid-IR resonances ($10-20$THz). An interesting way forward to modify heat load by cavities tuned below the mid-IR, is to combine them with optical filters that absorb the dominant mid-IR contribution to the Purcell mediated heat load. 

Importantly, the presence of such filters can be easily implemented by appropriately choosing the substrates of the cavity mirrors and our model indicates that the mirror substrates play an important role in the recent discovery of cavity control of phase transition. For a discussion on the role of cold filter in Ref.~\cite{Jarc2023} 
see appendix. In this case, the cavity modes can no longer be assumed to be at a thermal distribution as in the present study, and the choice of suitable filters becomes even more important in the prospect of controlling dissipations for the establishment on non-thermal steady states in materials \cite{FloresCalderon2023}.

 Finally, while in this work we focused on the discussion of quantum materials, the main conclusions are applicable and are expected to be relevant to a wide range of other settings such as for example molecules \cite{fassioli2021,nitzan2022}. Interestingly, in molecular systems typically characterized by sharp absorption lines (as opposed to the often broad absorption in quantum materials) heat flow may be dominant in a wide range of frequencies (and not necessarily the mid-IR), allowing for a heat control at ambient temperatures that is mediated by cavities at a wide range of frequencies without the need of optical filters.   Furthermore, a large enhancement of the temperature of only specific phonon modes could be achieved leading to non-thermal states.

\section*{Acknowledgements}
 This research is funded in part by the Gordon and Betty Moore Foundation, Grant GBMF12213 to DF.  
 ME acknowledges funding by the Deutsche Forschungsgemeinschaft (DFG, German Research Foundation), Project ID 429529648 – TRR 306 QuCoLiMa (“Quantum Cooperativity of Light and Matter”). 
 
 \section{Supplementary Information}
 \section{Electromagnetic density of states}

\subsection{General relations}

We assume the Hamiltonian for the empty cavity can be written in the form
\begin{align}
\label{Hexpan}
H_{\rm field}=\sum_{m} \hbar\omega_m \Big(\hat a_m^\dagger \hat a_m + \frac{1}{2}\Big),
\end{align}
where $a_m$ ($a_m^\dagger$) are annihilation (creation) operators for photons in the cavity normal modes $m$. Moreover, the (transverse) vector potential is expanded in normal modes as
\begin{align}
\label{Aexpan}
{\bm A}({\bm r}) = \sum_{m} \sqrt{\frac{\hbar}{2\epsilon_0 \omega_m V}} \big( {\bm \epsilon}_m({\bm r}) \hat a_m + h.c.\big),
\end{align}
where the mode functions ${\bm \epsilon}_m({\bm r})$ are ortho-normalized over the quantization volume $V$,
\begin{align}
\label{norm}
\int d^3r\, {\bm \epsilon}_m({\bm r})^* {\bm \epsilon}_n({\bm r}) = \delta_{n,m} V.
\end{align}
With ${\bm E} = - \partial_t {\bm A}$, the electric field is 
\begin{align}
\label{Eexpan}
{\bm E}({\bm r}) = \sum_{m} \sqrt{\frac{\hbar \omega_m }{2\epsilon_0  V}} \big( i{\bm \epsilon}_m({\bm r}) \hat a_m + h.c.\big).
\end{align}
For example, in free space  $m=({\bm k},\nu)$ are plane wave modes with wave vector ${\bm k}$ and two transverse polarizations $\nu=1,2$. The mode functions are ${\bm \epsilon}_{{\bm k},\nu}=\hat e_{{\bm k},\nu} e^{i{\bm k}{\bm r}}$, where $\hat e_{{\bm k},\nu}$ is the polarization vector ($\hat e_{{\bm k},\nu}\hat e_{{\bm k},\nu'}^*=\delta_{\nu,\nu'}$), and with the expansion \eqref{Aexpan} and \eqref{Eexpan}, the field energy \eqref{Hexpan}  is $H_{\rm field}=\frac{\epsilon_0}{2}\int d^3 r( \bm E^2 + c^2\bm B^2)$.

Of particular relevance below will be the partial local density of states for a given direction $\hat e$, which we define as 
\begin{align}
\label{doslocal}
N(\omega,{\bm r},\hat e) =\frac{1}{V} \sum_{m} |\hat e\cdot {\bm \epsilon}_m({\bm r})| \delta(\omega-\omega_m).
\end{align}
In free space, using the plane wave mode functions and $\omega_{\bm k,\nu}=ck$, one obtains $N(\omega,{\bm r},\hat e)=N_{\rm free}(\omega)$, with
\begin{align}
N_{\rm free}(\omega)=\frac{\omega^2}{3\pi^2c^3}
\end{align} 
being $1/3$ of the mode density of states $\frac{1}{V}\sum_{\bm k,\nu}\delta(\omega-ck)$. The local density of states is related to the (retarded) propagator of the electromagnetic field,
\begin{align}
\label{Daall}
D^R_{aa'}({\bm r},{\bm r}\,';t-t')= \frac{1}{i\hbar}\theta(t-t')
\big\langle [\hat E_a({\bm r},t),\hat E_{a'}({\bm r}\,',t')]\big\rangle_0.
\end{align}
Here $a=x,y,z$ denotes cartesian directions, and the time-dependence of the operators as well as the expectation value is understood with respect to the free Hamiltonian \eqref{Hexpan}, i.e., $\langle \cdots \rangle_0= Z_{0}^{-1}\text{tr} e^{-\beta H_{\rm field}}\cdots$. Using the expansion \eqref{Eexpan} within \eqref{Daall}, the Fourier transform $D^R(\omega+i0) = \int dt D^R(t)e^{i(\omega+i0)t} \equiv D'(\omega)+iD''(\omega)$ gives
\begin{align}
D^R_{aa'}&({\bm r},{\bm r}\,';\omega+i0)= 
\frac{1}{V}\sum_m
\frac{ \epsilon_{m,a}({\bm r}) \epsilon_{m,a'}({\bm r}\,')\omega_m^2}{\epsilon_0 [(\omega+i0)^2-\omega_m^2
]},
\end{align}
%\begin{align}
%D^R_{aa'}&({\bm r},{\bm r}\,';\omega+i0)= 
%\sum_m
%\frac{ \epsilon_{m,a}({\bm r}) \epsilon_{m,a'}({\bm r}\,')\omega_m}{2V\epsilon_0}
%\,\,\,\times
%\nonumber
%\\
%&\times\,\,\,\Big(\frac{1}{\omega+i0-\omega_m} - \frac{1}{\omega+i0+\omega_m} \Big),
%\end{align}
from which it follows that (for $\omega>0$)
\begin{align}
\label{NframD}
N(\omega,{\bm r},\hat e)=
-\frac{2\epsilon_0}{\omega\pi} 
%\sum_{a,a'} 
\hat e_a \hat e_{a'} D''_{aa'}({\bm r},{\bm r};\omega).
\end{align}
(Here and in the following we assume Einstein convention for the cartesian indices).

The local density of states is directly related to the radiation from a point dipole at position ${\bm r}$ with orientation $\hat e$: To show this, we write the interaction of the field with an external classical dipole source as  $H_{\rm ext}= \int d^3r \, p_a({\bm r},t) E_a({\bm r})$, where $p_a({\bm r},t)$ is a time-dependent polarization density. The radiative energy loss for an oscillating source $ p_a({\bm r},t)= \frac{1}{2}(p_a({\bm r})e^{-i\omega t} + c.c.)$ is given by the time-averaged rate of increase in the field energy, $W=\frac{1}{\tau}\int_0^{\tau} dt \langle dH_{\rm field}/dt \rangle$, $\tau=2\pi/\omega$, which in general is obtained from the imaginary part of the response function,
\begin{align}
W=- \frac{\omega}{2} \int d^3 r d^3r'  p_a({\bm r})p_b({\bm r}\,')D_{a,b}''({\bm r},{\bm r}',\omega).
\end{align}
Hence, for a point dipole at ${\bm r}_0$, $p_a({\bm r})= p_0\hat e\delta({\bm r}-{\bm r}_0)$ we obtain, using Eq.~\eqref{NframD},
\begin{align}
\label{pointDP}
W= p_0^2 \frac{\omega^2\pi}{4\epsilon_0} N(\omega, {\bm r}_0,\hat e).
\end{align}
For example, inserting the free space density of states, we obtain the standard result for the dipole radiation in free space, 
\begin{align}
W_0= p_0^2 \frac{\omega^4}{12\pi \epsilon_0 \pi c^3}.
\end{align}

\subsection{Electric field and density of states between two parallel mirrors}

Consider two parallel metallic mirrors  separated by a distance $L$. The z-axis is chosen to be normal to the planes of the mirrors, such that the mirrors are positioned at $z=0$ and $z=L$. Any position in space between the mirrors is characterized by ${\bm r} = z \hat{z} + \rho \hat{\rho}$, where $ \hat{\rho}$ is a vector parallel to the mirrors plane. It is convenient to also describe the wavevector in terms of components perpendicular and parallel to the mirrors surface as ${\bm k}=k_z \hat{z} + k_{\rho}\hat{\rho}$.
The presence of the mirrors, impose boundary conditions on the electric and magnetic fields at the mirrors surface where the in plane electric field ($E_{\rho}$) and  normal magnetic field ($B_z$) must vanish. This is satisfied by the condition on the vector potential $A_{\rho}(z)=0$ at $z=0$ and $z=L$ which impose a quantization of the z-component of the wavevector $k_z=n\pi/L$, with $n=0,1,2..$. The vector potential is then split into {\it transverse electric (TE)} and {\it transverse magnetic (TM)} contribution,
\begin{align}
{\bf A}({\bm r})  =  {\bf A}^{TE} ({\bm r}) + {\bf A}^{TM} ({\bm r})\label{A_vec}.
\end{align}
The mode functions are obtained by solving Maxwells equations with perfect mirrors as boundary conditions, imposing Eq.~\eqref{norm} as normalization condition. One obtains
\begin{align}
\label{epsTE}
\bm \epsilon^{\rm TE}_{k_{\rho},n}
 &=
  \hat{\rho} \times \hat{z}
\,\,  \sqrt{2}
%[\delta_{n,0} + 2^{1/2} (1-\delta_{n,0})] 
e^{i {k}_{\rho}\rho}
\sin \Big( \frac{n\pi z}{L}\Big)
\\
\bm \epsilon^{\rm TM}_{k_{\rho},n}
 &=
\frac{\sqrt{2}ck_{\rho}}{\omega_{k_{\rho},n}}
e^{i k_{\rho}\rho}
 \Big\{ 
 \frac{in\pi }{Lk_{\rho}} 
\sin \Big( \frac{n\pi z}{L} \Big) \hat{\rho} 
- 
\cos\Big(\frac{n\pi z}{L}\Big)\hat{z} \Big\}
\end{align}
 for $n=1,2,3,...$, and  an additional $n=0$ TM mode
\begin{align}
\label{epsTM0}
\bm \epsilon^{\rm TM}_{k_{\rho},0}
 &=
\sqrt{2}
e^{i k_{\rho}\rho}
 {\rm cos}(\frac{n\pi z}{L})\hat{z}.
 \end{align}
  The dispersion for both branches is 
\begin{align}
\omega_{k_{\rho},n}
=
\sqrt{\Big((n\omega_c)^2+ c^2k_{\rho}^2},
\end{align}
where  $\omega_c= \pi c/L$ is the fundamental frequency of a cavity mode propagating in the z-direction. Using $\bm E=-\partial_t \bm A$, and $\bm B=\bm\nabla\times \bm A$, one can see that $E_z=0$
and $B_z=0$ for the TE and TM modes, respectively.

Using the modes \eqref{epsTE} to \eqref{epsTM0} within Eq.~\eqref{doslocal}, we obtain the local density of states between two mirrors:
\begin{align}
\frac{N(\omega, z, \hat{\rho})}{N_{\rm free}(\omega)}
&=  \frac{3\omega_c}{2\omega} \sum_{n=1}^{\frac{\omega}{\omega_c} }
\Big[ 1 + \Big( \frac{n \omega_c}{\omega} \Big)^2 \Big] {\rm sin}^2 \Big( \frac{n\pi z}{L} \Big)
\end{align}
for polarization perpendicular to the mirrors, and  
\begin{align}
\frac{N(\omega,z,\hat{z})}{N_{\rm free}(\omega)}
&=    \frac{3\omega_c}{2\omega} +   \frac{3\omega_c}{\omega} \sum_{n=1}^{\frac{\omega}{\omega_c} } \Big[ 1 - \Big( \frac{n \omega_c}{\omega} \Big)^2 \Big] {\rm cos}^2 \Big( \frac{n\pi z}{L} \Big)
\end{align}
for polarization parallel to the mirrors.  

In the main text, we mainly need the density of states at the center between the two mirrors, $z=L/2$, where the sin and cos factors simply select the odd and even modes, respectively,
\begin{align}
\label{Dosparral}
\frac{N(\omega, L/2, \hat{\rho})}{N_{\rm free}(\omega)}
&=  \frac{3\omega_c}{2\omega} \sum_{n=1,3,...}^{\frac{\omega}{\omega_c} }
\Big[ 1 + \Big( \frac{n \omega_c}{\omega} \Big)^2 \Big] 
\\
\label{Dosperp}
\frac{N(\omega,L/2,\hat{z})}{N_{\rm free}(\omega)}
&=    \frac{3\omega_c}{2\omega} +   \frac{3\omega_c}{\omega} \sum_{n=2,4,...}^{\frac{\omega}{\omega_c} } \Big[ 1 - \Big( \frac{n \omega_c}{\omega} \Big)^2 \Big] .
\end{align}
(The sums can then be performed analytically.) The ratio $N/N_{\rm free}$ is shown in Fig.~1 of the main text. For the parallel polarization $\hat \rho$, one observes an enhancement and decrease just above/below the resonances $\omega=n\omega_c$ at odd $n$. In particular, Eq.~\eqref{Dosparral} gives $N(\omega, L/2, \hat{\rho})=0$ below the fundamental resonance $\omega=\omega_c$, and $N(\omega, L/2, \hat{\rho})=3N_{\rm free }(\omega)$ just above $\omega=\omega_c$. For the perpendicular polarization, the TM-0 mode gives rise to the first term $\frac{3\omega_c}{2\omega}$ in Eq.~\eqref{Dosperp}, which explains the divergence of the ratio $\frac{N(\omega,L/2,\hat{z})}{N_{\rm free}(\omega)}$ at $\omega\to 0$. Note, however, that $N(\omega,L/2,\hat{z})$ itself vanishes at $\omega\to0$, because $N_{\rm free}(\omega)\sim\omega^2$.  

We emphasise that the Purcell effect depends on the local density of states, which is very different from a spatial average.  The conclusion of Ref.~\cite{Chiriaco2023} that the thermal absorption between two coplanar mirrors is always reduced with respect to the free space was based a spatially averages density of states (also excluding the TM-0 mode). 

\subsection{Density of states between tilted mirrors}

In this subsection we compute the local density of states between two tilted mirrors, by computing the total radiation from a point dipole and Eq.~\eqref{pointDP}. We consider two perfect metallic plates (mirrors), with an opening angle $\alpha$. We choose the orientation such that the plates are spanned by vectors $\hat e_z$ and $\cos(\alpha/2)\hat e_x \pm  \sin(\alpha/2)\hat e_y$ and intersect at the line $x=y=0$. For simplicity we take $\alpha=\pi/N$ with integer $N$, which simplifies the solution in terms of image dipoles. A Hertz dipole $\bm p(t)=\bm p_0 e^{-i\omega t}$ is located at ${\bm r}_0= a \hat e_x$ with magnitude $p_0=|\bm p_0|$ and orientation $\hat p_0=\bm p_0/p_0 $ (without loss of generality, we take $p_0$ to be real). The radiation between the two plates is determined by the dipole and $2N-1$ image dipoles, at locations $\bm{r}_n = R_z(n\alpha){\bm r}_0$, where $R_z(\varphi)$ is the rotation matrix around the $\hat z$-axis, i.e.,
\begin{align}
\bm{r}_n = 
a\big(\cos(n\alpha) \hat e_x + \cos(n\alpha)\hat e_y\big).
\end{align}
Also the image dipoles are rotated, 
\begin{align}
\bm p_n =R_z(n\alpha)\, \big[(-1)^n(p_{0,x}\hat e_x + p_{0,z}\hat e_z) + p_{0,y}\hat e_y\big],
\end{align}
where the sign $(-1)^n$ takes into account a sign flip of the in plane components of the dipole at each mirror event. For a dipole $\bm p e^{-i\omega t}$ at the origin, the electric and magnetic field in the far field are given by
\begin{align}
\bm H_{\bm p} = i \frac{e^{-ikr}}{4\pi r} (\hat r\times \bm p) ck^2,
\,\,\,\,\,\bm E_{\bm p} = Z_0( \bm H \times \hat r),
\end{align}
where $Z_0=\sqrt{\mu_0/\epsilon_0}$, and $\hat r = {\bm r}/r$, and $k=\omega/c$. For dipole $n$ we thus obtain
\begin{align}
\bm H_{n} 
&= 
i \frac{e^{-ik|{\bm r}-{\bm r}_n|}}{4\pi |{\bm r}-{\bm r}_n|^2} ((|{\bm r}-{\bm r}_n)\times \bm p_n) ck^2
\\
&\sim
i \frac{e^{-ikr}}{4\pi r} e^{ik(\hat r {\bm r}_n)}(\hat r \times \bm p_n) ck^2.
\end{align}
In the second step we retained only the leading terms in the far field, using $|{\bm r}-{\bm r}_n| = r - (\hat r {\bm r}_n) + \mathcal{O}(1/r)$.
Hence the far field for the dipole and its images is given by
\begin{align}
\bm H 
&= 
i \frac{e^{-ikr}}{4\pi r}ck^2
\Big(\hat r \times \sum_{n=0}^{2N-1} \bm p_n e^{ik(\hat r {\bm r}_n)}\Big)
\\
&\equiv
i \frac{e^{-ikr}}{4\pi r}ck^2 p_0\,
\hat r \times \bm P(\hat r),
\end{align}
where  ($\hat p_0=\bm p_0/p_0$)
\begin{align}
\bm P
&
(\hat r,\hat p_0,ka)
= 
\sum_{n=0}^{2N-1} 
e^{ika[\hat r R_z(n\alpha)\hat e_x]} \,\,\,\times
\nonumber\\
&\times\,\,\,
R_z(n\alpha)\, \big[(-1)^n(\hat p_{0,x}\hat e_x +  \hat p_{0,z}\hat e_z) +  \hat p_{0,y}\hat e_y\big].
\end{align}
With $\bm E = Z_0( \bm H \times \hat r)$, the time-averaged pointing vector $\bm S = \frac{1}{2}\text{Re} \bm E^*\times \bm H$ is
\begin{align}
\bm S
%&= 
%p_0^2
%\frac{Z_0 k^4 c^2}{16\pi^2r^2}
%\frac{1}{2}\text{Re}
%\Big( ( (\hat r \times \bm P^*) \times \hat r)\times (\hat r \times \bm P) \Big)
%\\
%&= 
%\frac{Z_0 k^4 c^2}{16\pi^2r^2}
%\frac{1}{2}\text{Re}
%\Big(  (\hat r \times \bm P)\times (\hat r\times  (\hat r \times \bm P^*) ) \Big)
%\\
%&= 
%\frac{Z_0 k^4 c^2}{16\pi^2r^2}
%\frac{1}{2}\text{Re}
%\Big(  (\hat r \times \bm P)\times 
%(\hat r (\hat r \bm P^*) - \bm P^*)
%\Big)
%\\
%&= 
%\frac{Z_0 k^4 c^2}{16\pi^2r^2}
%\frac{1}{2}\text{Re}
%\Big(  
% (\hat r \bm P^*) 
%\hat r\times  ( \bm P \times \hat r)
%+ \bm P^* \times (\hat r \times \bm P)  )
%\Big)
%\\
&= 
\hat r
p_0^2
\frac{Z_0 k^4 c^2}{32\pi^2r^2}
\Big(   
 ( \bm P^* \bm P)  -(\hat r \bm P^*) ( \hat r\bm P )
\Big).
\end{align}
%where the second step follows from standard vector algebra.
We finally  (numerically) integrate the expression over all directions in which the dipole can radiate; in spherical coordinates $\hat r=\hat r(\vartheta,\varphi)$, the power is 
\begin{align}
W&= 
\int_0^\pi d\vartheta \sin(\vartheta) \int_{-\alpha/2}^{\alpha/2}
d\varphi
\, \hat r\bm S(\hat r),
\end{align}
and the density of states is obtained from Eq.~\eqref{pointDP}. The result depends only a function of $\alpha$, the dipole orientation $\hat p_0$,  and the product $ka$. 
Introducing $L(a)=2a\tan(\alpha/2)$ as the distance between the mirrors at the location of the dipoles, and $\omega_c(a)=c\pi/L(a)$ as the corresponding resonance frequency, we have
\begin{align}
ka  = \frac{\omega a}{c}
=
\frac{\omega}{\omega_c(a)} 
\frac{\pi}{2 \tan(\alpha/2)},
\end{align} 
which shows that $N/N_{\rm free}$ depends only on the ratio $\omega/\omega_c(a)$, the polarization, and the angle $\alpha$, as stated in the main text.

\section{Heat transfer between light and matter}

\subsection{Energy current}

In this section we derive the heat transfer between material and blackbody radiation under general assumptions, using the framework of linear response theory. We write the Hamiltonian in the form 
\begin{align}
\hat H = \hat H_{m} + \hat H_{\rm field} + \hat H_{I},
\end{align}
where $H_{m}$ and $H_{\rm field}$ are the Hamiltonian of field and matter, respectively, and $H_{I}$ is an  interaction
\begin{align}
\label{couling}
\hat H_{I} = \int d^3r \,\,\hat d_a  ({\bm r}) \hat E_a({\bm r}),
\end{align}
in the form of  a linear coupling between the electric field $\hat E({\bm r})$ and a polarization density $d({\bm r})$; both are treated as quantum mechanical operators at this point, $a$ is the cartesian component, and we again assume Einstein summation convention. 
(Note that the full light-matter interaction in the dipolar representation includes also a polarization self-interaction $\sim \int d^3r d({\bm r})^2$, which however is not relevant for calculating the heat transfer to leading order in the light-matter interaction.) The current density is $\hat J_a({\bm r}) = \frac{d}{dt} d_a ({\bm r}) =\dot d_a ({\bm r}) = i [H,d_{\alpha}({\bm r})] $. The energy current to the material is the time-derivative of the system energy due to the light-matter interaction,
\begin{align}
\dot Q =  i [\hat H,\hat H_m] =  i [\hat H_I,\hat H_m].
\end{align}
Using Eq.~\eqref{couling}, the energy current is therefore given by,
\begin{align}
\label{jj}
\dot Q = - \int d^3r  \hat J_a  ({\bm r}) \hat E_a({\bm r}).
\end{align}
We aim to calculate the energy transfer $ \langle \dot Q \rangle$, evaluated in the steady state. We will now proceed to express this heat current in terms of the response functions of the solid (polarizability) and the electromagnetic density of states.

\subsection{Field fluctuations}

The heat transfer does not only depend on the electromagnetic density of states, but also on the field fluctuations, which will be assumed to be thermal. Similar to the response function \eqref{Daall}, we introduce the autocorrelation functions,
\begin{align}
\label{EE1}
D_{a,a'}^>({\bm r},{\bm r}\,';t-t')&= -i\langle E_a({\bm r},t) E_{a'}({\bm r}\,',t')\rangle_0,\,\,\,
\\
\label{EE2}
D_{a,a'}^<({\bm r},{\bm r}\,';t-t')&= -i\langle  E_{a'}({\bm r}\,',t') E_a({\bm r},t)\rangle_0.
\end{align}
In a thermal state, the fluctuation dissipation theorem implies the universal relations 
for the Fourier transforms  ($D^>(\omega)=\int dt e^{i\omega t} D(t)$, etc.),
\begin{align}
\label{FDTEE1}
D_{a,a'}^>({\bm r},{\bm r}\,';\omega)&= 2i D''_{a,a'}({\bm r},{\bm r}\,';\omega)\big(1+b(\omega,T)\big),
\\
\label{FDTEE2}
D_{a,a'}^<({\bm r},{\bm r}\,';\omega)&= 2i D''_{a,a'}({\bm r},{\bm r}\,';\omega)\,b(\omega,T),
\end{align}
where $b(\omega,T)$ is the Bose function. The relation is proven by expanding $D^R$, $D^<$, and $D^>$ in an eigenbasis of $H_{\rm field}$.

\subsection{Response functions of the solid}

The polarizability of the solid (i.e., the response of $d$ to an external field) is given by the Kubo response relation
\begin{align}
\label{Kubo}
\alpha^R_{a,a'}({\bm r},{\bm r}\,';t-t')= -\frac{i}{\hbar} \theta(t-t') \langle [\hat d_a({\bm r},t), \hat d_{a'}({\bm r}\,',t')] \rangle_0.
\end{align}
Analogous to the field correlations, we can also introduce the autocorrelation functions
\begin{align}
\label{dd1}
\alpha_{a,a'}^>({\bm r},{\bm r}\,';t-t')&= -i\langle d_a({\bm r},t) d_{a'}({\bm r}\,',t')\rangle_0,\,\,\,
\\
\label{dd2}
\alpha_{a,a'}^<({\bm r},{\bm r}\,';t-t')&= -i\langle  d_{a'}({\bm r}\,',t') d_a({\bm r},t)\rangle_0.
\end{align}
Again, in  a thermal state, the fluctuation dissipation theorem implies the universal relations  for the Fourier transforms  
\begin{align}
\label{FDTdd1}
\alpha_{a,a'}^>({\bm r},{\bm r}\,';\omega)&= 2i\alpha''_{a,a'}({\bm r},{\bm r}\,';\omega)\big(1+b(\omega,T)\big),
\\
\label{FDTdd2}
\alpha_{a,a'}^<({\bm r},{\bm r}\,';\omega)&= 2i\alpha''_{a,a'}({\bm r},{\bm r}\,';\omega)\,b(\omega,T),
\end{align}
where we again use the notation $\alpha^R(\omega+i0)=\alpha'(\omega)+i\alpha''(\omega)$.
The above expressions still depend on ${\bm r}$ and therefore represent the full momentum-dependent polarizability. Later we will later need to consider only the long wavelength $(\vec q=0)$ limit.

\subsection{Heat flow}

To compute the heat flow, work in the limit of weak system bath coupling, and use the Kubo formula to compute the expectation value $\langle \dot Q\rangle $ due to the steady perturbation $H_{I}$,
\begin{align}
\langle\dot Q\rangle 
&= \frac{1}{i\hbar}\int_{-\infty}^t dt' \langle [\dot Q(t),\hat H_I(t')] \rangle_{0} e^{0^+t}
\\
&=
\frac{1}{\hbar}\text{Im} \int_{0}^\infty dt \langle [\dot Q(t),\hat H_I(0)] \rangle_{0} e^{-0^+t}
\end{align}
Here the expectation value and time dependence of operators is now taken in the equilibrium state of decoupled light of matter, and in the second equality we have used translational invariance as well as the fact that $\langle\dot Q\rangle$ is real. Inserting \eqref{couling} and \eqref{jj},
\begin{align}
\langle \dot Q \rangle 
&= 
-\frac{1}{\hbar}\text{Im}
\int_{0}^\infty dt 
\int d^3rd^3r'
\nonumber\\
&\,\,\,\,\langle [\hat J_a  ({\bm r},t) \hat E_a({\bm r},t),\hat d_b ({\bm r}\,',0) \hat E_b({\bm r}\,',0)] \rangle_{0}.
\end{align}
Now light and matter expectation values can be factorized, leading to
\begin{align}
\langle \dot Q \rangle 
= 
&-\frac{1}{\hbar}\text{Im}
\int_{0}^\infty dt 
\int d^3rd^3r'
\nonumber\\
&
\Big(
\langle \hat J_a  ({\bm r},t)  \hat d_b ({\bm r}\,',0)\rangle_{0} \langle\hat E_a({\bm r},t) \hat E_b({\bm r}\,',0)\rangle_{0}
\nonumber
\\
&-
\langle \hat d_b ({\bm r}\,',0) \hat J_a  ({\bm r},t)  \rangle_{0} \langle\hat E_b({\bm r}\,',0)\hat E_a({\bm r},t) \rangle_{0}
\Big).
\end{align}
Using $J=\partial_t d$, the heat flow can now be expressed in the correlation functions \eqref{EE1}, \eqref{EE2}, \eqref{dd1}, and \eqref{dd2}.
\begin{align}
\langle \dot Q \rangle 
= 
&\hbar\,\text{Im}
\int_{0}^\infty dt 
\int d^3rd^3r'
\Big(
D_{a,b}^>({\bm r},{\bm r}\,';t)
 \partial_t 
 \alpha_{b,a}^<({\bm r}\,',{\bm r};-t) 
\nonumber\\
&-
D_{a,b}^<({\bm r},{\bm r}\,';t)
\partial_t 
\alpha^>_{b,a}({\bm r}\,',{\bm r};-t)
\Big).
\end{align}
Transforming to frequency representation,
\begin{align}
\langle \dot Q \rangle 
= 
&\int \frac{d\omega}{4\pi}
\int d^3rd^3r'
\hbar\omega
\Big(
D_{a,b}^>({\bm r},{\bm r}\,';\omega)
 \alpha_{b,a}^<({\bm r}\,',{\bm r};\omega) 
\nonumber\\
&-
D_{a,b}^<({\bm r},{\bm r}\,';\omega)
\alpha^>_{b,a}({\bm r}\,',{\bm r};\omega)
\Big).
\end{align}
We now assume that both light and matter are in separate thermal quasi-equilibrium at temperatures $T_l$  and $T_m$, respectively. Hence we can use the fluctuation dissipation relations \eqref{FDTEE1}, \eqref{FDTEE2}, \eqref{FDTdd1}, and \eqref{FDTdd2}.
\begin{align}
\label{efeghss}
\langle \dot Q \rangle 
&= 
\int \frac{d\omega}{\pi}
\int d^3rd^3r'
D_{a,b}''({\bm r},{\bm r}\,';\omega)
\alpha''_{b,a}({\bm r}\,',{\bm r};\omega)
\,\,\,\,\,\times
\nonumber\\
&\times\,\,\,\,
\hbar\omega
\Big(
b(\omega,T_l)
-
b(\omega,T_m)
\Big).
\end{align}
We note that the factor $\omega (b(\omega,T_l)-b(\omega,T_m))$ is $>0$ ($<0$) if $T_l>T_m$ ($T_m>T_l$), i.e., heat is correctly flowing from the hot to the cold reservoir. To further simplify Eq.~\eqref{efeghss}, we note that the bose functions restrict the frequency integral to the range of THz to GHz frequencies at room temperature.  For the cavity geometries of interest here (coplanar or tilted cavity, the field correlation function is therefore varying on scales much slower than the atomic distance).  Hence we can make the local density approximation, which corresponds to replacing 
\begin{align}
\alpha''_{b,a}({\bm r}\,',{\bm r};\omega)
\approx 
\delta({\bm r}\,' - {\bm r}) \alpha''_{b,a}(\omega),
\end{align}
where $\alpha''_{b,a}(\omega) = \frac{1}{V} \int d^3rd^3r'\alpha''_{b,a}({\bm r}\,',{\bm r};\omega)$ is the long wavelength ($q=0$) limit of the polarizability. Moreover, we use that the integrand is symmetric with respect to $\omega$. Hence we have, assuming for simplicity that the material is homogeneous,
\begin{align}
\label{LDA efeghss}
\langle \dot Q \rangle 
&= 
2\int_0^\infty \frac{d\omega}{\pi}
\int d^3r 
D_{a,b}''({\bm r},{\bm r};\omega)
\alpha''_{b,a}(\omega)
\,\,\,\,\,\times
\nonumber\\
&\times\,\,\,\,
\hbar\omega
\Big(
b(\omega,T_l)
-
b(\omega,T_m)
\Big).
\end{align}
For simplicity, let us assume the principle axis of the polarizability tensor are $x,y,z$. Then the product 
$D_{a,b}''  \alpha''_{b,a}$ reduces to $\sum_a D_{a,a}''  \alpha''_{a,a} $, and 
\begin{align}
\label{LDA efeghss1}
\langle \dot Q \rangle 
&= 
\sum_{a}
\int_0^\infty d\omega
\int d^3r 
\frac{\hbar\pi\omega^2}{\epsilon_0}N(\omega,{\bm r},\hat e_a)
\,\,\,\,\,\times
\nonumber\\
&\times\,\,\,\,
\Big(-\frac{1}{\pi}\alpha''_{a,a}(\omega)\Big)
\Big(
b(\omega,T_l)
-
b(\omega,T_m)
\Big).
\end{align}
where Eq.~\eqref{NframD} was used to reintroduce the local density of states. 
In the main text, we express the result in terms of the conductivity instead of the polarizability, using the general relation
%
%  It may be more convenient to express the result in terms of a conductivity. The conducvity (response of current to field $E$) is given by a Kubo relation similar to Eq.~\eqref{Kubo},
%\begin{align}
%\sigma^R_{a,a'}({\bm r},{\bm r}\,';t-t')= -\frac{i}{\hbar} \theta(t-t') \langle [\hat J_a({\bm r},t), \hat d_{a'}({\bm r}\,',t')] \rangle_0,
%\end{align}
%and since $J=\partial_t d$, we have 
%\begin{align}
%\sigma^R_{a,a'}(\omega)= -i\omega \alpha^R_{a,a'}(\omega),
%\end{align}
%i.e., 
$\alpha''(\omega) = -\frac{ \sigma'(\omega)}{\omega}$. With this we obtain Eq.~(1) of the main text,
\begin{align}
\label{LDA efeghss4}
\langle \dot Q \rangle 
&= 
\sum_{a}
\int_0^\infty d\omega
\int d^3r 
\frac{1}{\epsilon_0}N(\omega,{\bm r},\hat e_a)
\sigma_{aa}'(\omega)
\,\,\,\,\,\times
\nonumber\\
&\times\,\,\,\,
\hbar\omega
\Big(
b(\omega,T_l)
-
b(\omega,T_m)
\Big).
\end{align}

\begin{figure}
\centerline{\includegraphics[width=0.99\columnwidth]{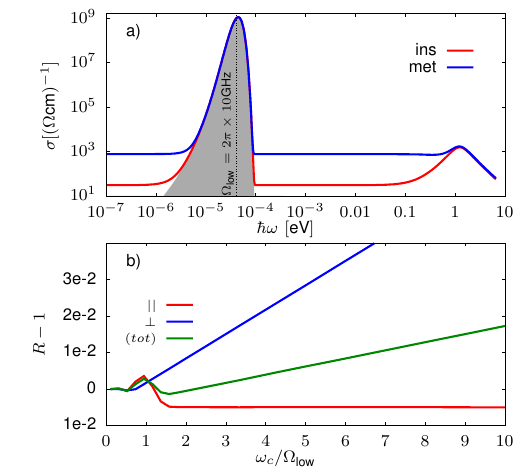}}
%\centerline{\includegraphics[width=0.99\columnwidth]{plots/fig04_suggestion.png}}
\caption{
a) Solid lines: Model conductivity of 1T-TaS$_2$  with a sum two Lorentz functions \eqref{lorenz_appendix} and a low frequency peak at $\Omega_{\rm low}=2\pi\times 10$GHz (shaded region, see text). Parameters for Lorenz oscillators are $(\Omega,\omega_p,\gamma) =(1,3,1)\text{~and~}(0,1.5,0.5)$  for the metallic phase, and  $(\Omega,\omega_p,\gamma) =(1,3,1)\text{~and~}(0,0.3,0.5)$ for the insulating phase, in units of $2\pi\times 10^4$cm$^{-1}$.  b) The ratio R between heat flow in cavity and free space [Eq.~(4) in main text] for the insulating model, $T_l=300$K, $T_m=150$K, and  cavity frequencies around  $\omega_c=\Omega_{\rm low}. $The different lines show a cases with conductivity purely in plane ($||$), out of place ($\perp$), and isotropic (tot).
}
\label{fig5}
\end{figure}

\section{Discussion for {\em 1T}-${\rm \bf TaS}_2$ }  

We now discuss the implications 
%ME01 
our %of the above 
analysis 
for the experimental setting in \cite{Jarc2023}. The experiment observes a large cavity induced modification of the apparent phase transition temperature. In a thermal scenario as proposed in the manuscript, this would imply a large modification of the heat flow between photon bath and material, i.e., a large modification of the ratio $R=\dot q/\dot q_{\rm free}$ defined in Eq.~(4) of the main text.  If the material under investigation ($1T$-TaS$_2$) would conduct in the THz range, the analysis of the main text (Fig.~2) suggests that the thermal Purcell effect would be negligible for the setting of Ref.~\cite{Jarc2023}, where cavities with $\omega_c<1$THz were employed.  However, the conductivity of 1T-TaS$_2$ has been reported in the THz range \cite{Gasparov2002}, but, to the best of our knowledge, it is not know for frequencies in the frequency range of the cavity resonance [10-50GHz]. Here we therefore examine whether an unknown strong low-frequency contribution to the conductivity can explain the observed behavior, as proposed in Ref.~\cite{Jarc2023}.

First, one could assume that the low frequency conductivity is dominated by a single Lorentz oscillator
\begin{align}
\label{lorenz_appendix}
\sigma_{\Omega,\gamma,\omega_p}(\omega) &=  \epsilon_0\frac{\omega_p^2 \gamma \omega^2}{(\Omega^2-\omega^2)^2 + \omega^2 \gamma^2},
\end{align}
%ME NOTE the 2pi factor!
at  frequency of $10$GHz  ($\Omega\approx 2\pi\times10$ GHz), corresponding to the relevant cavity frequency, and a damping $\gamma\sim\Omega$ or larger. In this case, the photon temperature would be much larger than $\Omega$, ($T_l\approx 300$K $ \gg \hbar\Omega/k_B=0.5$K). As discussed in connection with Fig.~3 of the main text, the contribution of the GHz-oscillator to the heat transfer would therefore still be dominated by the high-frequency tails of the Lorentzian (up $\omega\sim T_l$), and the cavity at $\omega_c\ll T_l$ would have little effect of the heat transfer.

%ME01 ... optional?
However, a real system will not display a perfect Lorentz from of the absorption peak even at large detuning from the resonance. 
To avoid the effect from these high frequency tails, one could assume a more localized form for the conductivity in the GHz range, such as a Gaussian 
%ME added a factor (omega/Omega)^2 to have a low frequency behavior ~ omega^2 (like the Loretzntzian). Quantitatively, the factor not very significant
 $\sigma_{\rm GHz}\times (\omega/\Omega)^2 e^{-(\omega-\Omega)^2/\Omega^2}$, which can arise from an inhomogeneous distribution of narrow oscillators.  If also the high-frequency THz conductivity is contributing to the heat flow, this ansatz would however require a very large value for the conductivity $\sigma_{\rm GHz}$ in order for this low frequency band to be significant to the overall heat flow:   For example, one can compare the contribution to the heat flow of a flat conductivity $\sigma_{flat}=1$A.U. (where heat flow is maximized in the THz range) with that of a low frequency Gaussian conductivity $\sigma_{GHz}=1$ A.U., taking again $\Omega=2\pi\times 10$GHz, $T_l=300$K, and $T_m=150$K, one obtains  a ratio $Q_{\rm GHz}/Q_{\rm flat}=10^{-10}$ between the heat flow from the low frequency band and the flat background. In turn, this would indicate that  the amplitude $\sigma_{\rm GHz}$  would have to be orders of magnitude larger than the background conductivity, for the GHz cavity to have a sizable effect on the heat flow.
 
 A more detailed illustration is given in Fig.~\ref{fig5}: Here we model the conductivity of 1T-TaS$_2$ with two Lorentz functions, describing the DC conductivity (Drude peak) and the excitation across the gap, respectively \cite{Gasparov2002}. In addition, we add a strong Gaussian peak at a frequency $\Omega_{\rm low}=2\pi\times10$GHz. The resulting ratio $R$ between the heat flow in the cavity and in free space [Eq.~(4) of the main text] is shown in Fig.~\ref{fig5}b) for the insulating phase. Consistent with the estimate above, one finds that, in order for the cavity to have an effect of few per cent on $R$, the amplitude $\sigma_{\rm GHz}$ of the Gaussian should be $8$ orders of magnitude larger than the background DC conductivity. While the conductivity in the GHz peak is not known, a contribution of this magnitude would certainly have a measurable effect on the inductive response at higher frequencies, via the Kramers Kronig 
 %ME01
relation, and appears unrealistic. % relation and appears unrealistic. 
\\
 
 On the other hand in the previous discussion a perfect mirror is assumed (i.e. a mirror that allows all wavelengths to be equally reflected and transmitted). In realistic cavities, the transmission will strongly depend not only on the metallic mirrors used, but also on the transmission of the substrate of such mirrors. Furthermore, the sample is held in between membranes, both of which can in principle filter out specific wavelengths. 
 Indeed, in \cite{Jarc2023} the quartz substrate of the gold cavity mirrors used in the experiment, maximize transmission in the GHz range and inhibits it at the higher  mid-IR frequencies. This effectively acts as a spectral filter which could reduce the, otherwise dominant mid-IR contribution to the total heat load of the sample. This reduction could allow the thermal Purcell effect due to GHz cavity to modify the sample temperature. 
 %For example, if we assume a perfect filter of the radiation above $0.1$THz, the ratio of the heat flow due to a flat conductivity $\sigma_{flat}=1$A.U. %and  a low frequency Gaussian conductivity  with $\sigma_{GHz}=1$ A.U ($\Omega=2\pi\times 10$GHz) is increased from the estimate $Q_{\rm GHz}/Q_{\rm %flat}\approx10^{-10}$ to $Q_{\rm GHz}/Q_{\rm  flat}\approx10^{-3}$. Correspondingly, an effect of the cavity would already be measurable for a moderate GHz peak (about $7$ orders of magnitude smaller than in Fig.~\ref{fig5}). 
 Crucially, if the mirrors or membranes are absorbing, the cavity modes will no longer be in a thermal distribution which may have important implications on whether the formation of the states in matter are thermal or not.  The above discussion shows that a proper design of the filtering properties of the whole setup is essential to control the radiative heat loads in real experiments and a study in this direction is left for future works.

\bibliography{references}% Produces the bibliography via BibTeX.

\end{document}